\documentclass[aps,twocolumn,groupedaddress]{revtex4}
\usepackage{amsmath}
\usepackage{graphics}

\begin{document}
\bibliographystyle{revtex}
\preprint{TP-CB-371}

\title{Probing Pauli Blocking Factors in Quantum Pumps with Broken Time-Reversal Symmetry}

\author{Mathias Wagner}
\email[]{wagner@phy.cam.ac.uk}
\affiliation{Hitachi Cambridge Laboratory, Madingley Road,
             Cambridge CB3 0HE, United Kingdom}

\date{\today}

\begin{abstract}
A recently demonstrated quantum electron pump is discussed within the 
framework of photon-assisted tunneling. Due to lack of time-reversal 
symmetry, different results are obtained for the pump current depending 
on whether or not final-state Pauli blocking factors are used when  
describing the tunneling process. Whilst in both cases the current 
depends quadratically on the driving amplitude for moderate pumping, a 
marked difference is predicted for the temperature dependence. With 
blocking factors the pump current decreases roughly linearly with 
temperature until $k_{\rm B} T$ $\approx$ $\hbar\omega$ is reached, 
whereas without them it is unaffected by temperature, indicating that 
the entire Fermi sea participates in the electronic transport.
\end{abstract}

\maketitle

\begin{figure}[b]
  \centering
  \includegraphics{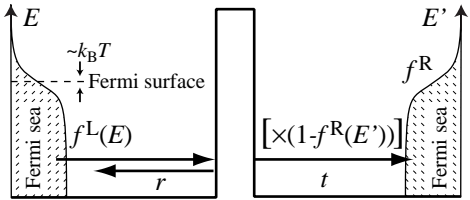}
   \caption{Single-barrier scattering state with transmitted and 
   reflected amplitudes, populated according to a distribution 
   function $f^{\rm L}$ characterising the left side. In the Pauli 
   picture, the tunneling probability incurs an additional 
   final-state blocking factor $1-f^{\rm R}$ on the right.
}
\label{Single-Barrier}
\end{figure}

Tunneling of electrons through classically forbidden regions is one of 
the major paradigms of quantum mechanics. Although our understanding 
has greatly improved over the past decades, some vital aspects of this 
process are still subject to fierce debates. For example, no consensus 
has been reached at yet whether so-called final-state blocking factors 
exist in the tunneling process across a barrier sandwiched between two 
conductors (Fig.~\ref{Single-Barrier}), to enforce Pauli's exclusion 
principle that no two electrons may occupy the same quantum state. 
This seemingly innocent question is related to the question where in 
the so-called Fermi sea of conducting electrons the current 
actually flows, and thus has far-reaching consequences for our 
understanding of quantum transport. There are two schools on how to 
calculate the tunnel current, one that insists on using the blocking 
factors, and another that rejects them \cite{Datta95a}. The dilemma is 
that both schools almost always seem to give identical answers for the 
current. In this Letter we point out that  this result is 
fundamentally related to time-reversal symmetry, and study a generic
system where this symmetry is broken to gain insight into the nature 
of blocking factors.

Consider a tunneling barrier impeding the current flow as depicted in 
Fig.~\ref{Single-Barrier}. According to the first school, we start 
by calculating the {\it coherent} scattering states of the system 
consisting of an incident plane wave, a reflected wave, and a 
transmitted wave on the far side. These scattering states travel either 
from left to right or the opposite way. In the stationary limit they 
can simply be populated according to distribution functions 
$f^{\rm L}$ and $f^{\rm R}$ governing the asymptotic contact regions 
on either side. This yields for the current \cite{Datta95a}
\begin{eqnarray}
    I_{\rm S} & = & 
 {2e\over h} \int dE dE' \,{\cal D}_\bot T^+(E',E) f^{\rm L}(E) \cr
               & - &
 {2e\over h} \int dE dE' \,{\cal D}_\bot T^-(E,E') f^{\rm R}(E') ,
    \label{eq:Tsu-Esaki-Current}
\end{eqnarray}
where  ${\cal D}_\bot$ is a density-of-states factor \cite{footnote1},
$T^+(E',E)$ is the transmission probability for scattering states 
incident from the left at energy $E$ and emerging at the right at $E'$ 
($\ne E$ in general), and $T^-$ is defined in a similar manner for the 
reverse direction. According to Eq.~(\ref{eq:Tsu-Esaki-Current}) 
{\it all} scattering states contribute to the current, even those far
below the Fermi surface --- as recently discussed for the Fermi pump 
\cite{Wagner99a}. It is only when the transmission probabilities are 
symmetric, $T^+(E',E)$ = $T^-(E',E)$, that the {\it net} current seems 
to stem from electrons in the immediate vicinity of the Fermi surface 
only, as in this case all electrons further down cancel each other. 

An alternative and widely used recipe for calculating the tunneling 
current in Fig.~\ref{Single-Barrier} is based on the 
transfer-Hamiltonian formalism, originally put forward by Oppenheimer 
\cite{Oppenheimer28a}, and later refined by Bardeen and others 
\cite{Duke69a}. Here the system is split into two subsystems, the 
left- and the right-hand side, with a common overlap in the central 
barrier region, and the \ result for the current can be expressed as
\begin{eqnarray}
    I_{\rm P} & = & 
 {2e\over h} \int dE dE'  \,{\cal D}_\bot T^+(E',E) 
	                       [1-f^{\rm R}(E')]f^{\rm L}(E) \cr
              & - &
 {2e\over h} \int dE dE'  \,{\cal D}_\bot T^-(E,E') 
	                       [1-f^{\rm L}(E)]f^{\rm R}(E')  
    \label{eq:Pauli-Current}
\end{eqnarray}
which differs from Eq.~(\ref{eq:Tsu-Esaki-Current}) by the Pauli 
blocking factors $1-f$. These factors are introduced --- more or less 
{\it ad hoc} --- using the intuitive argument that an electron 
tunneling from one side to the other needs to find an empty final 
state on the far side to tunnel into \cite{Sturman84a}. 

Clearly, the philosophies underlying Eqs.~(\ref{eq:Tsu-Esaki-Current}) 
and (\ref{eq:Pauli-Current}) are entirely different: For the former we 
assume that the left- and right-hand side of the barrier belong to the 
{\it same} system and that the scattering state extends coherently 
across the barrier, whereas in the latter we consider {\it real} 
incoherent transitions between initial and final states on opposite 
sides of the barrier belonging to {\it different} subsystems
\cite{footnote2}.
Both views have their merits. Certainly in the limit of a very 
transparent or even vanishing barrier, one has to regard both sides of 
the barrier as part of the same system, in which case there is no place 
for blocking factors. But equally well, if tunneling is weak, one can 
argue that the two sides do form two different subsystems, and that 
final-state blocking factors are mandatory to guarantee Pauli's 
exclusion principle and to prevent an ``overflow'' of the final state. 

Crucially, this analysis also applies to a system with {\it local} 
time-dependent driving forces as its leads, and hence the distribution 
of incident electrons, are not affected by the driving field. For a 
system with time-reversal symmetry, it can be shown that the 
transmission probabilities are such 
that any channel $E \leftrightarrow E'$ is traversed with the same 
probability in either direction, i.e. that microreversibility holds, 
$T^+(E',E)$ $\equiv$ $T^-(E,E')$. Under this condition the cross terms 
$f^{\rm L}f^{\rm R}$ arising from the Pauli blocking factors cancel and 
Eqs.~(\ref{eq:Tsu-Esaki-Current}) and (\ref{eq:Pauli-Current}) yield 
identical answers for the {\it total} current, which is the quantity 
usually measured in experiment. It is probably because of this 
indecisive result that both schools have coexisted for so long. The 
only difference between the two is their prediction {\it where} in 
the Fermi sea the current flows: With blocking factors, the current is 
forced to flow close to the Fermi surface, as in this picture 
lower-lying final states are blocked, whilst in the formulation based 
on scattering states the current flows in the entire Fermi sea. 

However, if time-reversal symmetry does not hold,  
Eqs.~(\ref{eq:Tsu-Esaki-Current}) and (\ref{eq:Pauli-Current}) yield 
different answers even for the {\it magnitude} of the current itself 
as the cross terms do not cancel any longer. This is very intriguing 
as it gives us hope to come closer to an experimentally verifiable 
answer regarding the existence of Pauli blocking factors. 
\begin{figure}[bt]
  \centering
  \includegraphics{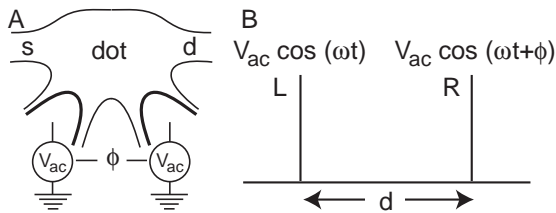}
   \caption{Quantum-dot pump with broken time-reversal symmetry. 
   [A] Schematic drawing; [B] simple 1D  model based on driven 
   $\delta$-function barriers.
}
\label{Pump}
\end{figure}

Our analysis is inspired by a recent experiment by the Marcus group 
\cite{Switkes99a}. A semiconductor quantum dot with source and drain 
point contacts has two additional lateral gates to which ac voltages 
of relative phase $\phi$ are applied (see Fig.~\ref{Pump} A). 
Time-reversal symmetry is broken unless $\phi$ is an integer multiple 
of $\pi$. Averaging over different dot configurations, they measured 
fluctuations of the emf voltage generated between the source and drain 
contacts. Previous theoretical studies employed the concept of 
adiabatic pumping \cite{Altshuler99a}.
We take a complementary perspective and view the pump current (which 
exists even in the {\it absence} of any applied source-drain bias) as 
due to photon-assisted tunneling \cite{Wagner95a}. 

As a model system of a driven dot strongly coupled to its leads we 
consider two harmonically oscillating $\delta$-function barriers of 
equal strength $V_{\rm ac}$ a distance $d$ apart in a 1D potential, 
with a variable phase difference $\phi$ in the ac signals 
as depicted in Fig.~\ref{Pump} B \cite{Hekking91a}. This is a very 
simplified model of the Marcus pump, but nevertheless it turns out to 
exhibit many of its characteristic properties. For calculating the 
transmission probabilities $T(E',E)$ across the dot we take advantage 
of the fact that we can split this problem into two parts: If we know 
the transmission and reflection amplitudes for each barrier separately, 
we can use the Fabry-Perot method of raytracing known in optics to 
calculate the interference pattern due to multiple reflections between 
the barriers. As in optics the partial interference amplitudes can be 
summed up to all orders in a geometric series, yielding for the 
transmission amplitudes at the far side \cite{Wagner97b}
\begin{eqnarray}
   t & = & t^R (I - K)^{-1} Q t^L ,
   \label{Fabry-Perot} 
\end{eqnarray}
where $K$ = $Qr^LQr^R$ describes one full round trip of the electron 
between the two barriers $L$ and $R$, starting and ending at the $R$ 
barrier. Each round trip an electron at energy $E_n$ = $E$ + 
$n\hbar\omega$ picks up a phase factor $Q_n$ = $\exp[i(k_n d+\theta)]$ 
consisting of two parts: The phase $k_n d$ incurred after travelling a 
distance $d$ with wave vector $k_n$ = $\sqrt{2mE_n}/\hbar$, and a 
fixed but unknown phase $\theta$. This latter phase is introduced to 
account for the random changes in magnetic field and dot 
geometry employed in the experiment to perform ensemble averages. In 
general, the phase $\theta$ depends on the electron energy, but for 
simplicity we ignore this and assume $\theta$ to be equally 
distributed. Under this condition ensemble averaging simply means 
averaging over $\theta$ at the end of the calculation. 

Due to the discreteness of the photon energy, electrons emerge on the 
far side of the barriers at energies differing from their 
original energy $E$ by multiples of $\hbar\omega$: $E'$ = $E$ + 
$n\hbar\omega$, the so-called sidebands. Consequently, all amplitudes 
in Eq.~(\ref{Fabry-Perot}) have to be interpreted as  
{\it matrices}, and the transmission probability takes the form 
$T(E',E)$ = $\sum_n$ $\tilde T_n(E) \delta(E+n\hbar\omega-E')$.

Before presenting numerical results for strong driving, it is 
instructive to study the weak-driving limit, which can be solved 
analytically. Defining $T_{\rm e}(k,\varphi)$ = $\eta^2$ $(k_0/ 4 k)$ 
$\cos^2\left[d(k_0 + k)/2 + \varphi/2\right]$ where $\eta$ = 
$V_{\rm ac}k_0/E_0$ is the dimensionless pump amplitude and $k_0$ the 
wave vector of the incident electron, we obtain for the transmission 
probabilities in the first three sidebands up to $\eta^2$ 
\cite{footnote3} 
\begin{eqnarray}
  \tilde T^+_0(\phi,\theta) & = & 
    1 + T_{\rm e}(-k_1,-\phi) + T_{\rm e}(-k_{-1},\phi) \cr
    & &
    - T_{\rm e}(k_1,2\theta-\phi) - T_{\rm e}(k_{-1},2\theta+\phi) \cr
  \tilde T^+_{\pm 1}(\phi,\theta) & = & - T_{\rm e}(-k_{\pm 1},\mp \phi) .
    \label{eq:T01}
\end{eqnarray}
The corresponding transmission probabilities $\tilde T^-_n$ for the 
reverse direction are obtained by substituting $(\phi,\theta)$ $\to$ 
$(-\phi,\theta)$. As expected, the net transmission probability 
$\tilde T_{\rm net}$ $\equiv$ $\tilde T^+$ - $\tilde T^-$ vanishes for 
$\phi$ = $n\pi$ when time-reversal symmetry holds, and is maximal at 
$\phi$ = $\pi/2$. In agreement with experimental findings, the pump 
current at zero dc bias, being proportional to $\tilde T_{\rm net}$, 
scales with $\eta^2$ for weak driving. In the experiment the photon 
energy is 6 orders of magnitude smaller than the Fermi energy. 
Expanding Eq.~(\ref{eq:T01}) in this limit the current turns out to 
be {\it linear} in the driving frequency, in perfect 
agreement with the experimental results available \cite{footnote4}.
\begin{figure}[tb]
  \centering
  \includegraphics{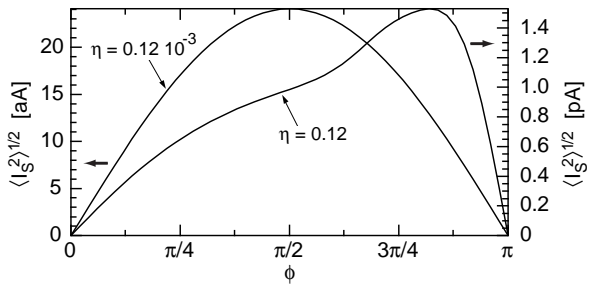}
  \caption{Dependence of $\langle I_{\rm S}^2 \rangle^{1/2}$ on the 
      phase shift $\phi$ for weak and strong driving fields. 
	Parameters: $d$ = 0.2\,$\mu$m, $f$ = 10\,MHz, $m$ = 0.067\,$m_0$, 
	$E_F$ = 12\,meV, $T$ = 0.1\,K.
}
\label{I2-Dependence-On-Phi}
\end{figure}

\begin{figure}[tb]
  \centering
  \includegraphics{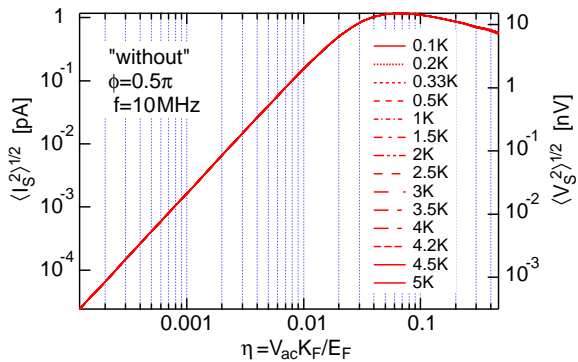}
  \caption{$\langle I_{\rm S}^2(\pi/2) \rangle^{1/2}$
	   and $\langle V_{\rm S}^2(\pi/2) \rangle^{1/2}$,
	   both based on Eq.~(\protect\ref{eq:Tsu-Esaki-Current}),
	   as a function of the driving strength $V_{\rm ac}$ for 
	   a range of temperatures.
}
\label{Dependence-On-Vac-TE}
\end{figure}
After integrating over $\theta$ the ensemble average 
$\langle\tilde T_{\rm net}\rangle$ is found to be identically zero up 
to order $\eta^2$. This implies that the direct current 
$\langle I \rangle$ is orders of magnitude smaller than its fluctuations 
for all but the strongest driving fields, and hence we will concentrate 
on the mean square average $\langle I^2 \rangle^{1/2}$ to study the 
fluctuations instead \cite{footnote5}.
Figure~\ref{I2-Dependence-On-Phi} shows the dependence of 
$\langle I_{\rm S}^2 \rangle^{1/2}$ on the phase shift $\phi$, where 
we have used Eq.~(\ref{Fabry-Perot}) with an adaptive number of photon 
sidebands to determine the full non-linear transmission probability, 
and Eq.~(\ref{eq:Tsu-Esaki-Current}) to finally calculate the current. 
For small driving amplitudes we find a $\sin\phi$ behaviour in 
agreement with Eq.~(\ref{eq:T01}), whilst for stronger driving 
nonharmonic features appear, both of which are in qualitative 
agreement with experiment. A similar behaviour is observed when using 
the formula (\ref{eq:Pauli-Current}) based on Pauli blocking factors 
instead, except that in this case the pump current is much smaller.

For the remainder of this Letter we will fix the ac phase shift $\phi$ 
to $\pi/2$, the point of maximal time-reversal asymmetry. The current
fluctuations $\langle I_{\rm S}^2(\pi/2) \rangle^{1/2}$ calculated
using the formula (\ref{eq:Tsu-Esaki-Current}) {\it without} blocking 
factors are illustrated in Fig.~\ref{Dependence-On-Vac-TE} as a 
function of $\eta$, taken at the Fermi surface. For small driving 
amplitudes up to $\eta$ $\approx$ 0.05 the fluctuations increase 
quadratically with $\eta$ as suggested by Eq.~(\ref{eq:T01}) before 
eventually starting to decrease. In the Marcus experiment, rather than 
measuring the pump current, the emf voltage generated was studied. 
Defining the emf voltage as the difference in chemical potentials 
between the left and right leads necessary to make the pump current 
vanish, we find that its fluctuations, also shown in 
Fig.~\ref{Dependence-On-Vac-TE} (right-hand axis), have virtually the 
same dependence on $\eta$ as the current fluctuations. Both the 
quadratic rise as well as the levelling off for stronger pumping have 
been observed in experiment. The maximal emf fluctuations generated, 
15\,nV, is only one or two orders smaller than in the experiment, 
which is a quite reasonable agreement given the simplicity of the 
model.
\begin{figure}[tb]
  \centering
  \includegraphics{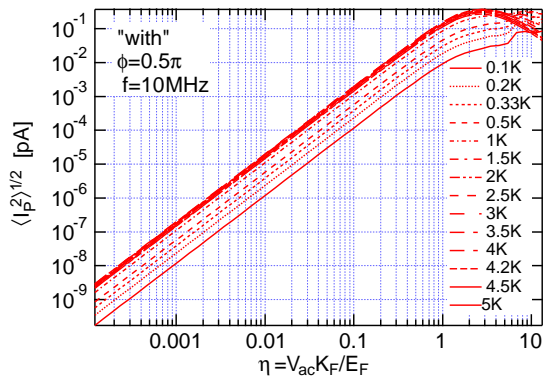}
  \caption{$\langle I_{\rm P}^2(\pi/2) \rangle^{1/2}$, based on 
         Eq.~(\protect\ref{eq:Pauli-Current}), as a function of the 
         driving strength $V_{\rm ac}$ for a range of temperatures.
}
\label{Dependence-On-Vac-P}
\end{figure}

The corresponding results based on the alternative formula 
(\ref{eq:Pauli-Current}) for the current which does include the Pauli 
blocking factors are illustrated in Fig.~\ref{Dependence-On-Vac-P}. 
Similar to the case without blocking factors of 
Fig.~\ref{Dependence-On-Vac-TE}, the current fluctuations 
$\langle I_{\rm P}^2(\pi/2) \rangle^{1/2}$ (as well as the emf 
fluctuations, not shown) display an $\eta^2$ behaviour first before 
eventually starting to saturate. Yet, not only do the magnitudes of 
these results differ substantially from Fig.~\ref{Dependence-On-Vac-TE} 
for the same driving amplitude, but now there is a very pronounced 
temperature dependence as well!

\begin{figure}[b]
  \centering
  \includegraphics{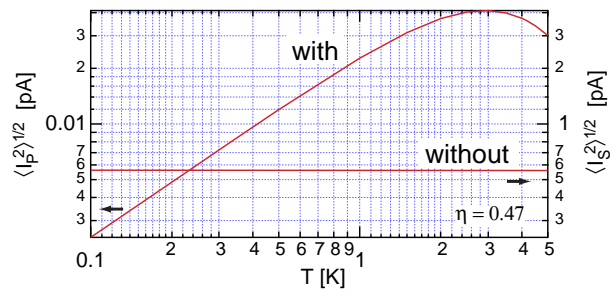}
  \vspace{-0.2cm}
  \caption{Temperature dependence of $\langle I^2(\pi/2) \rangle^{1/2}$
  with and without Pauli blocking factors.
}
\label{Dependence-On-T}
\end{figure}

This temperature dependence is worked out in more detail in
Fig.~\ref{Dependence-On-T} using a fixed driving strength of $\eta$ = 
0.47. Whilst without Pauli blocking factors the variation with 
temperature is minimal, an almost linear dependence is observed when 
they are included. This distinctly different behaviour in the 
low-temperature regime is most easily understood when looking at the 
number of states in phase space effectively available for transport: 
The blocking factors force the current to flow within a few 
$k_{\rm B} T$ of the Fermi surface (see Fig.~\ref{Single-Barrier}), 
and as the temperature approaches zero, this range of active 
current-carrying states eventually diminishes to a minimal width of a 
few $\hbar\omega$, which in the experiment is much smaller than 
$k_{\rm B} T$. However, since each state can only carry a certain 
maximal load, it follows that the pump current must also decrease with 
temperature, until it settles for a residual value once $k_{\rm B} T$ 
$\approx$ $\hbar\omega$ is reached --- with our parameters at 
$\approx$ 0.5\,mK. On the other hand, without blocking factors there 
are no phase-space restrictions, in which case the pump current flows 
in the entire Fermi sea and thus is largely immune to changes at the 
Fermi surface brought about by temperature. 

In the Marcus experiment the pump current is found to {\it increase} when 
lowering the temperature, and appears to level off at $\approx$ 0.1\,K, 
where phase-breaking events become less important \cite{Switkes99a}. 
Such events are not included in our theory, and we can therefore not 
expect to reproduce the high-temperature behaviour. However, the 
experimental finding of a {\it saturated} pump current in the 
low-temperature limit is (if genuine and not due to thermal decoupling) 
only consistent with our results based on the scattering-state approach, 
Eq.~(\ref{eq:Tsu-Esaki-Current}), but {\it not} with the formulation 
relying on Pauli blocking factors, i.e. Eq.~(\ref{eq:Pauli-Current}).

Being able to prove the (non)existence of Pauli blocking factors has 
drastic consequences for deciding whether the current flows in the 
entire Fermi sea or at its surface only. In this Letter we have 
demonstrated that a powerful tool for studying this issue is to look 
at the temperature dependence of the pump current when breaking 
time-reversal symmetry. Although our model system is simple, the 
conclusions about the low-temperature behaviour, being drawn from 
phase-space considerations, are clearly of a much more universal 
nature.

The author acknowledges discussions with C.~Marcus, and
support by the EU (FMRX-CT98-0180).

\bibliography{PauliFactors,DrivenSystems}

\begin{thebibliography}{10}
\bibitem{footnote1}
Depending on the energy $E_\bot$ $\equiv$ $E'_\bot$ 
perpendicular to the tunneling, which we assume to be conserved.
\bibitem{footnote2}
This is the same reasoning as in Fermi's Golden Rule, and so by analogy 
the transfer-Hamiltonian formalism only applies for small transmission 
probabilities.
\bibitem{footnote4}
This linear relationship has been said to be a fingerprint of 
adiabatic pumping, but as seen here photon-assisted tunneling 
yields the same dependency.
\bibitem{footnote5}
To avoid spurious low-energy resonance effects in the 
dot (which would not exist had we included potential fluctuations) 
we impose a lower cutoff to the energy integration in 
Eqs.~(\protect\ref{eq:Tsu-Esaki-Current}) and 
(\protect\ref{eq:Pauli-Current}) at 1\,GHz. Changing this cutoff leads 
to small quantitative but no qualitative changes in our results.
\end{thebibliography}

\end{document}